\documentclass{aa}

\usepackage{amsmath}
\usepackage{amssymb}
\usepackage{setspace}
\usepackage{multirow}

\usepackage[version=3]{mhchem}   
\usepackage[nottoc, notlof, notlot]{tocbibind}   
\usepackage{enumitem}
\usepackage{graphicx}
\usepackage{color}
\usepackage{colortbl}
\usepackage{txfonts}

\usepackage{hyperref} 
\bibpunct{(}{)}{;}{a}{}{,}             

\begin{document}

\title{Complex organic molecules in diffuse clouds along the line of sight to Sgr B2}
\author{V. Thiel\inst{1} \and A. Belloche\inst{1} \and K. M. Menten\inst{1} \and R. T. Garrod\inst{2} \and H. S. P. M\"uller\inst{3}}
  \institute{Max-Planck-Institut f\"ur Radioastronomie, Auf dem H\"ugel 69, 53121 Bonn, Germany \and Departments of Chemistry and Astronomy, University of Virginia, Charlottesville, VA 22904, USA \and I. Physikalisches Institut, Universit\"at zu K\"oln, Z\"ulpicher Str. 77, 50937 K\"oln, Germany}


\abstract
{Up to now, mostly relatively simple molecules have been detected in interstellar diffuse molecular clouds in our galaxy, but more complex species have been reported in the diffuse/translucent medium of a $z = 0.89$ spiral galaxy.}{We aim at searching for complex organic molecules (COMs) in diffuse molecular clouds along the line of sight to Sgr\,B2(N), taking advantage of the high sensitivity and angular resolution of the Atacama Large Millimeter/submillimeter Array (ALMA). }{We use data acquired as part of the EMoCA survey performed with ALMA. To analyse the absorption features of the molecules detected towards the ultracompact \ion{H}{ii} region K4 in Sgr~B2(N), we calculate synthetic spectra for these molecules and fit their column densities, line widths, centroid velocities, and excitation temperatures.}{We report the detection of CH$_3$OH, CH$_3$CN, CH$_3$CHO, HC$_3$N, and NH$_2$CHO in Galactic center (GC) diffuse clouds and CH$_3$OH and CH$_3$CN in a diffuse cloud in the Scutum arm. The chemical composition of one of the diffuse GC clouds is found to be similar to the one of the diffuse/translucent medium of the $z=0.89$ spiral galaxy.}{The chemical processes leading to chemical complexity in the diffuse molecular ISM appear to have remained similar since $z=0.89$. As proposed in previous studies, the presence of COMs in diffuse molecular clouds may result from a cyclical interstellar process of cloud contraction and expansion between diffuse and dense states.}


\maketitle

\section{Introduction}
About 200 different molecules have been detected up to now in the interstellar medium (ISM)\footnote{see, e.g. https://www.astro.uni-koeln.de/cdms/molecules}. Carbon-bearing molecules with at least six atoms are commonly referred to as complex organic molecules (COMs). They are detected in dense environments such as cold prestellar cores and hot cores or corinos \citep{herbst2009}. They are usually detected in emission, but several have been detected in absorption, for instance in the envelope of Sgr~B2 \citep{corby2015}. COMs have been detected not only in such dense environments, but also in photodissociation regions (PDRs) like the Horsehead \citep{guzman2014}. Here, the strong far-ultraviolet (FUV) radiation seems to play an important role: not only can the COMs survive in such environments, but the UV field also seems to enhance their abundance compared to neighbouring dense clouds \citep{guzman2014}. In addition, several COMs have been detected in absorption in a $z=0.89$ spiral galaxy, with a chemical composition of this absorber suggesting that the medium is diffuse/transculent \citep{muller2011,muller2014}. Thus, the question arises as to whether COMs also exist in galactic diffuse clouds, in which dust extinction is modest and densities are low. Up to now, mostly fairly simple molecules, such as CN, HCN, CCH, HCO$^+$ or c-C$_3$H$_2$ , have been detected in diffuse molecular clouds in our galaxy \citep[e.g.][]{lucas1997,godard2010}.
 
The low densities in diffuse molecular clouds result in low excitation temperatures, close to the temperature of the cosmic microwave background (CMB) radiation, that is, 2.73 K \citep{greaves1992}. Under these conditions, rotational lines are subthermally excited, very weak, and difficult to detect in emission. Absorption studies have a better sensitivity, but a strong continuum background source is needed. The giant molecular cloud Sagittarius B2 (Sgr\,B2) fulfills this condition. It is located near the Galactic centre (GC) at a projected distance of 100\,pc and a distance of $8.34\pm0.16$\,kpc to the sun \citep{reid2014}.

For our analysis, we used the EMoCA survey that aims at exploring the chemical complexity of the interstellar medium \citep{belloche2016}. This survey was performed towards the star forming region Sgr B2(N) with the Atacama Large Millimeter/submillimeter Array (ALMA) in Cycles 0 and 1. The spatial resolution of this survey is high enough to resolve the structure of Sgr\,B2(N) (see Fig.~\ref{cont}a). In this way, we can investigate absorption lines at positions where the continuum is still strong enough but which are sufficiently far away from the hot cores towards which absorption features are blended with numerous emission lines. The ultracompact \ion{H}{ii} region K4 \citep{gaume1995} fulfills these requirements. With the high sensitivity of the EMoCA survey we can search for absorption from COMs along the whole 8\,kpc-long line of sight to the Galactic centre.  

\section{Observations and analysis method}\label{sect_obs}
\begin{figure*}
\centering
\includegraphics[width=17cm, trim = 1.9cm 7.8cm 4.2cm 2.8cm, clip=True]{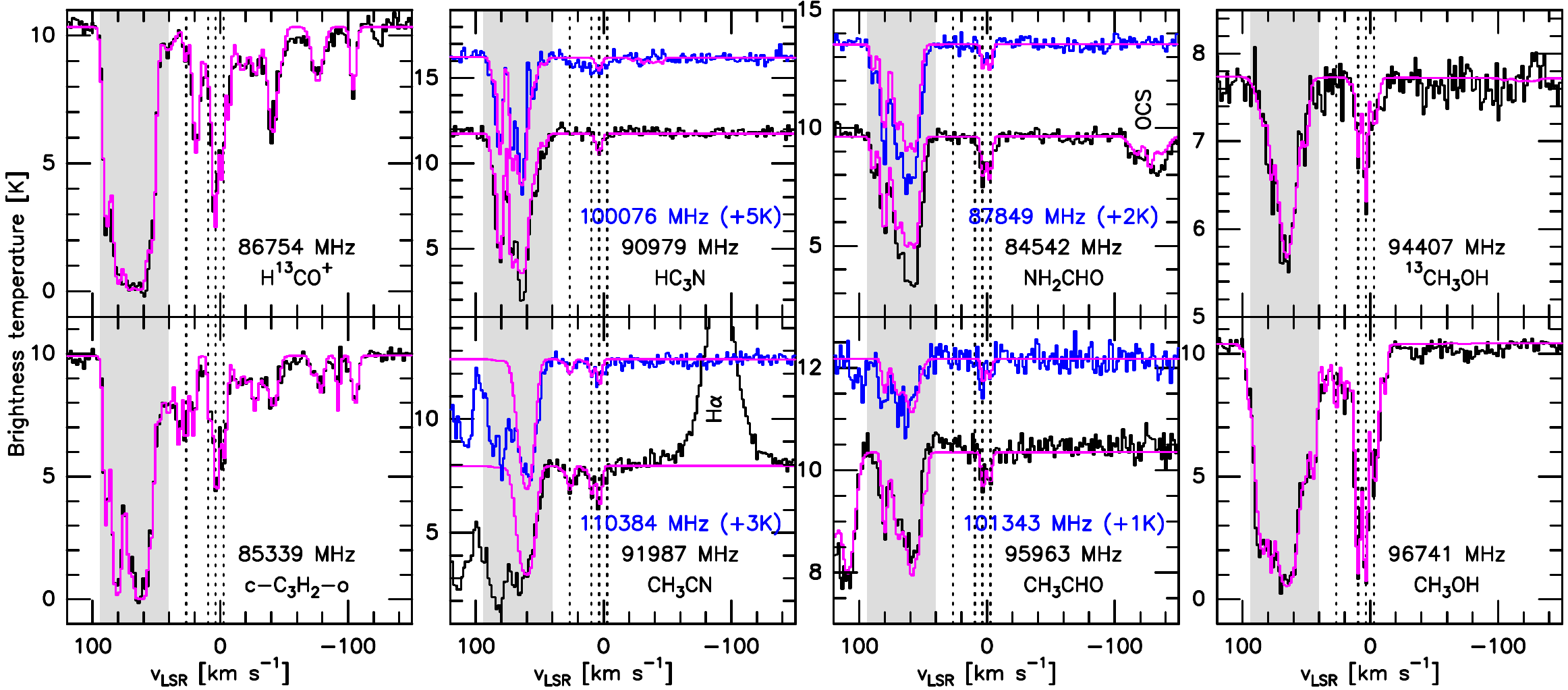}
\caption{Absorption spectra of simple and complex organic molecules towards the UC\ion{H}{ii} region K4 in Sgr~B2(N). The strongest transition is shown in black and the next strongest, if any, in blue. The molecules and rest frequencies of the transitions are given in each panel. The blue spectra are shifted upwards by the value given in parentheses. The synthetic spectra are overlain in magenta. The four dotted lines mark the median centroid velocities of the components detected for the complex species: 26.7, 9.4, 3.4, and $-2.8$~km\,s$^{-1}$. The absorption line at 110~km\,s$^{-1}$ close to the line of CH$_3$CHO at 95963~MHz is another transition of the same molecule. The grey area indicates the velocity range of the Sgr\,B2 envelope.}
\label{spectra}
\end{figure*}
We use the EMoCA survey taken with ALMA towards Sgr\,B2(N). The phase centre (EQ\,J2000: $17^\mathrm{h}47^\mathrm{m}19.87^\mathrm{s},-28^\circ$22\arcmin 16\arcsec) is half way between the two main hot cores N1 and N2 (see Fig.~\ref{cont}a). The survey covers the frequency range from 84.1 to 114.4 GHz with a median angular resolution of $1\farcs6$. The spectral resolution is 488~kHz or 1.3--1.7~km\,s$^{-1}$. The average noise level is \textasciitilde 3~mJy\,beam$^{-1}$. The calibration and deconvolution of the data were performed by \citet{belloche2016}. We corrected the spectra for the primary beam attenuation. We use Weeds \citep{maret2011} to model the spectra. We assume that all transitions of a molecule have the same excitation temperature and that the beam filling factor is 1. We fit the column density, linewidth, and centroid velocity. We assume the excitation temperature to be equal to the temperature of the CMB (2.73\,K) except in the cases where several transitions are detected, for which the temperature is derived from a population diagram.

\section{Results}\label{sect_res}
To maximise the sensitivity, we searched for absorption lines of COMs towards the ultracompact \ion{H}{ii} region K4, which is the strongest continuum source in the field of view except for the hot cores (see Fig.~\ref{cont}a). This source is located at ($\alpha$,$\delta$)$_\mathrm{J2000}=(17^\mathrm{h}47^\mathrm{m}20.02^\mathrm{s},-28^\circ$22\arcmin 04.7\arcsec), $13.8\arcsec$ to the north of N1. Many absorption lines are detected towards this position. In total, we identified lines from 19 different molecules in the diffuse molecular clouds along the line of sight to the GC: CN, CCH, c-C$_3$H$_2$, HNC, HCN, HCO$^+$, SiO, SO, CH$_3$OH, CO, CS, N$_2$H$^+$, HOC$^+$, HNCO, H$_2$CS, CH$_3$CN, CH$_3$CHO, HC$_3$N, and NH$_2$CHO. For some of these molecules we detected several isotopologues. For the identification we used the CDMS database \citep{mueller2005,mueller2001} and the JPL catalogue \citep{pickett1998}. In addition, we detected C$_2$H$_5$OH, CH$_3$SH, and CH$_3$NH$_2$ in absorption at velocities corresponding to the envelope of Sgr\,B2.

Here, we focus on the molecules with at least five atoms detected in the diffuse clouds: c-C$_3$H$_2$-o (ortho), CH$_3$OH, CH$_3$CN, CH$_3$CHO, HC$_3$N, and NH$_2$CHO. For comparison purposes, we also present results for H$^{13}$CO$^+$. The transitions investigated in this work are listed in Table~\ref{transitions}. Their spectra are shown in Fig.~\ref{spectra}. Apart from c-C$_3$H$_2$ and H$^{13}$CO$^+$ that are detected in more clouds, we detect absorption of these molecules at velocities corresponding to diffuse clouds in the GC and, for CH$_3$OH and CH$_3$CN, also in the Scutum arm. These components (at 26.7, 9.4, 3.4, and $-2.8$~km\,s$^{-1}$) are marked with dotted lines in Fig.~\ref{spectra}. The strong absorption components at $\sim$64~km\,s$^{-1}$ and $\sim$80~km\,s$^{-1}$ belong to the envelope of Sgr\,B2 and will not be analysed here. The velocity components at around 9.4, 3.4, and $-2.8$~km\,s$^{-1}$ fall roughly in the range of Galactic centre clouds and the component at 26.7~km\,s$^{-1}$ lies in the range of the Scutum arm \citep[e.g.][and references therein]{indriolo2015,menten2011}. Figure~\ref{cont}b shows as an example the intensity map integrated over the velocity range of the GC diffuse clouds in one of the transitions of NH$_2$CHO. The absorption is detected towards K4 only and the spatial extent of the absorption cannot be measured with our data set due to a lack of sensitivity.

The results of the modelling are given in Table~\ref{gc_clouds} for the GC clouds and Table~\ref{scut_arm} for the Scutum arm. To calculate the upper limits for $^{13}$CH$_3$OH we assumed the same velocity and linewidth as for CH$_3$OH. To determine the upper limits for the other non-detected molecules we took the median values of the centroid velocities and linewidths as fixed values. The linewidths range between 3 and 6\,km\,s$^{-1}$ for the GC centre clouds and between 3.5 and 5\,km\,s$^{-1}$ for the Scutum arm cloud. They fall in the same range as found in other investigations of diffuse molecular clouds along the line of sight to Sgr\,B2 \citep[e.g.][]{menten2011}. There are no systematic kinematical differences between the simple molecules H$^{13}$CO$^+$ and c-C$_3$H$_2$ and the more complex molecules, so they probably all trace the same regions.

\begin{table*}
\caption{Model parameters for the three diffuse GC clouds.}
\vspace*{-0.3cm}
\label{gc_clouds}
\centering
\setlength{\tabcolsep}{1.7mm}
\begin{tabular}{l r c c c r c c c r c c c}       
\hline              
Molecule & \multicolumn{1}{c}{$N_\mathrm{tot}$} & $v_\mathrm{LSR}$ & $FWHM$ & & \multicolumn{1}{c}{$N_\mathrm{tot}$} & $v_\mathrm{LSR}$ & $FWHM$ & & \multicolumn{1}{c}{$N_\mathrm{tot}$} & $v_\mathrm{LSR}$ & $FWHM$ & $T_\mathrm{ex}$ \\
 & \multicolumn{1}{c}{[cm$^{-2}$]} & [km\,s$^{-1}$] & [km\,s$^{-1}$] & & \multicolumn{1}{c}{[cm$^{-2}$]} & [km\,s$^{-1}$] & [km\,s$^{-1}$] & & \multicolumn{1}{c}{[cm$^{-2}$]} & [km\,s$^{-1}$] & [km\,s$^{-1}$] & [K] \\
\hline\hline
H$^{13}$CO$^+$ & 1.5$\times 10^{12}$ & 9.8 & 5.5 & & 8.0$\times 10^{12}$ & 3.8 & 4.5 && 3.8$\times 10^{12}$ & $-1.7$ & 3.5 & 2.73 (*) \\
c-C$_3$H$_2$-o & 5.0$\times 10^{12}$ & 8.2 & 3.0 && 1.9$\times 10^{13}$ & 2.8 & 4.5 && 1.2$\times 10^{13}$ & $-3.2$ & 4.5 & 2.73 (*) \\   
CH$_3$OH & 3.8$\times 10^{14}$ & 9.4 & 4.0 && 3.8$\times 10^{14}$ & 3.7 & 3.0 && 2.1$\times 10^{14}$ & $-3.6$ & 6.0 & 2.73 (*) \\
$^{13}$CH$_3$OH & 2.0$\times 10^{13}$ & 9.4 & 4.0 && 2.9$\times 10^{13}$ & 3.7 & 4.0 && 2.0$\times 10^{13}$ & $-3.6$ & 6.0 & 2.73 (*) \\ 
CH$_3$CN & 1.0$\times 10^{13}$ & 8.7 & 4.0 && 2.0$\times 10^{13}$ & 3.0 & 4.0 && <6.0$\times 10^{13}$ & $-2.8$ & 4.0 & $4.2\pm0.7$ \\
CH$_3$CHO & <1.5$\times 10^{13}$ & 9.4 & 4.0 && 5.0$\times 10^{13}$ & 2.6 & 3.5 && 4.0$\times 10^{13}$ & $-2.4$ & 3.5 &  $3.1\pm0.1$ \\
HC$_3$N & <2.5$\times 10^{13}$ & 9.4 & 4.0 && 6.0$\times 10^{14}$ & 3.7 & 5.0 && <2.5$\times 10^{13}$     & $-2.8$ & 4.0 & $7.3\pm0.8$ \\
NH$_2$CHO & <4.0$\times 10^{12}$ & 9.4 & 4.0 && 1.7$\times 10^{13}$ & 2.6 & 3.5 && 1.8$\times 10^{13}$ & $-2.4$ & 3.5 & $4.3\pm0.2$ \\
\hline                      
\end{tabular}
\vspace*{-0.35cm}
\tablefoot{The excitation temperatures marked with a star are assumed, the other ones were fitted.}
\end{table*}

We plot in Fig.~\ref{ntots}a the column densities derived for the diffuse clouds and compare them to the column densities in the $z=0.89$ spiral galaxy in front of the QSO PKS~1830--211 \citep[][and S. Muller, priv. comm.]{muller2011, muller2013}, the Horsehead PDR \citep[][and V. Guzm\'an, priv comm.]{pety2005, goicoechea2009, gratier2013, guzman2013, guzman2014}, the hot core Sgr\,B2(N2) \citep{belloche2017,belloche2016, mueller2016}, and the envelope of the Class 0 protostar NGC\,1333-IRAS 2A \citep{taquet2015}. The HCO$^+$ column densities for the diffuse clouds investigated here are derived from the H$^{13}$CO$^+$ column densities assuming an isotopic ratio $^{12}$C/$^{13}$C equal to 20 for the GC and 40 for the Scutum arm \citep{milam2005}.

\begin{table}
\caption{Model parameters for the diffuse cloud in the Scutum arm.}
\vspace*{-0.3cm}
\label{scut_arm}
\centering
\begin{tabular}{l r c c c}       
\hline               
Molecule & \multicolumn{1}{c}{$N_\mathrm{tot}$} & $v_\mathrm{LSR}$ & $FWHM$ & $T_\mathrm{ex}$ \\
 & \multicolumn{1}{c}{[cm$^{-2}$]} & [km\,s$^{-1}$] & [km\,s$^{-1}$] & [K] \\
\hline\hline              
H$^{13}$CO$^+$ & 6.2$\times10^{11}$ & 27.0 & 3.5 & 2.73\\
c-C$_3$H$_2$-o & 8.1$\times10^{12}$ & 26.8 & 3.5 & 2.73\\
CH$_3$OH & 6.2$\times10^{13}$ & 26.6 & 5.0 & 2.73\\
$^{13}$CH$_3$OH & <3.9$\times10^{12}$ & 26.6 & 5.0 & 2.73\\
CH$_3$CN & 1.4$\times10^{13}$ & 26.2 & 5.0      & $4.2$ \\
CH$_3$CHO & <1.5$\times10^{13}$ & 26.7 & 4.3 & $3.1$\\
HC$_3$N & <2.5$\times10^{13}$ & 26.7 & 4.3 & $7.3$\\
NH$_2$CHO & <4.0$\times10^{12}$ & 26.7 & 4.3 & $4.3$\\
\hline                      
\end{tabular}
\vspace*{-0.25cm}
\tablefoot{We assumed the same excitation temperatures as in Table~\ref{gc_clouds}.}
\end{table}

The velocity component at around 3.4~km\,s$^{-1}$ is relatively strong for all molecules. The one at $-2.8$~km\,s$^{-1}$ has comparable column densities, except for HC$_3$N and CH$_3$CN that are not detected. The component at 9.4~km\,s$^{-1}$ is in general weaker than the previous ones, except for methanol and methyl cyanide for which it is nearly as strong as the first component. CH$_3$CN in the Scutum arm cloud (26.7~km\,s$^{-1}$) has a similar column density as in the GC clouds, but CH$_3$OH is weaker. 

The \ion{H}{i} and H$_2$ column densities of the GC and Scutum-arm diffuse clouds are also plotted in Fig.~\ref{ntots}a (see also Table~\ref{hi_h2}). They were calculated by integrating the column density distributions of \citet{winkel2017} over velocity ranges equal to the median $FWHM$ of the absorption components in this work. The \ion{H}{i} data have an angular resolution of $\sim 37\arcsec$ \citep{winkel2017} and the HF data used to determine the H$_2$ column densities $\sim 40\arcsec$ \citep{bergin2010}. The \ion{H}{i} and H$_2$ column densities in the GC clouds are up to one order of magnitude higher than in the Scutum arm. The molecular fraction is 0.70 for the Scutum-arm cloud and 0.4--0.6 for the GC clouds (Table~\ref{hi_h2}).

Figure~\ref{ntots}b shows the abundances relative to methanol. In addition to the sources shown in Fig.~\ref{ntots}a, the abundances in the translucent cloud CB24 \citep{turner1998,turner1998a,turner1999} are also plotted. They correspond to the mean values (and standard deviations as uncertainties) of the different models presented by these authors. Apart from HC$_3$N, the chemical composition relative to methanol of the 3.4~km~s$^{-1}$ GC diffuse cloud is similar to the one of the $z=0.89$ spiral galaxy. But no obvious similarity between the COM compositions of these two sources and the other types of sources is apparent in Fig.~\ref{ntots}. The diffuse GC clouds, the translucent cloud CB24, the hot core Sgr~B2(N2), and the $z=0.89$ spiral galaxy have similar [CH$_3$CN]/[CH$_3$OH] ratios while the Scutum arm cloud and the Horsehead PDR lie one order of magnitude above and the protostellar envelope one order of magnitude below. The diffuse GC clouds, the $z=0.89$ absorber, and Sgr~B2(N2) have similar [NH$_2$CHO]/[CH$_3$OH] ratios while the protostellar envelope is nearly two orders of magnitude below. The [HC$_3$N]/[CH$_3$OH] ratio shows a wide spread with the 3.4~km~s$^{-1}$ GC component lying four orders of magnitude above the protostellar envelope, and the Horsehead PDR, Sgr~B2(N2), CB24, and the $z=0.89$ spiral galaxy being at a similar intermediate level. Finally all sources except for Sgr~B2(N2) have similar [CH$_3$CHO]/[CH$_3$OH] ratios.

\section{Discussion}\label{sect_disc}

\begin{figure*}
\centering
\includegraphics[width=17cm, trim = 1.5cm 0.3cm 1.5cm 1.cm, clip=True]{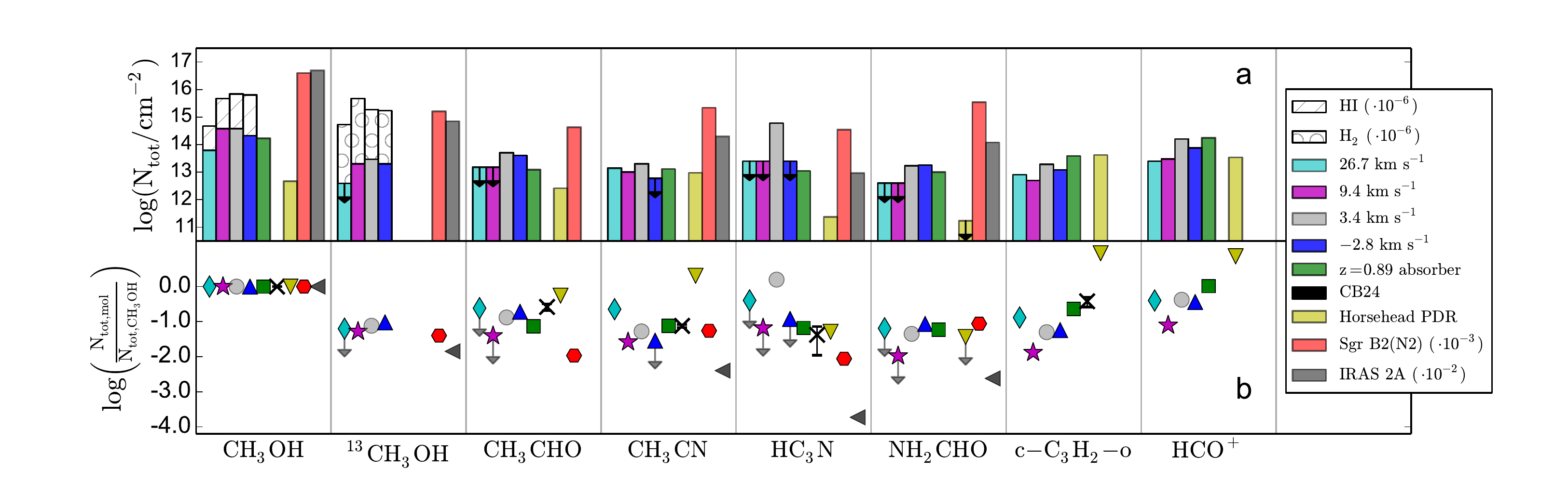}
\caption{\textbf{a} Column densities of complex organic molecules and two simpler ones detected in diffuse clouds in the GC and in the Scutum arm compared to the column densities in the $z=0.89$ spiral galaxy in front of PKS~1830-211, the Horsehead PDR, the hot core Sgr\,B2(N2), and the protostellar envelope of NGC\,1333-IRAS\,2A  (see references in Sect.~\ref{sect_res}). The column densities of \ion{H}{i} and H$_2$ in the diffuse clouds are shown in the panels of methanol \citep{winkel2017}. The values for \ion{H}{i}, H$_2$, the hot core, and the protostellar envelope are scaled by the factor given in parentheses. \textbf{b} Abundances relative to CH$_3$OH. The black crosses show the ratios for the translucent cloud CB24 \citep{turner1998,turner1998a,turner1999}. The arrows indicate upper limits.}
\label{ntots}
\end{figure*}

All complex molecules reported in Sect.~\ref{sect_res} are detected in the diffuse GC cloud at $\sim$3.4~km\,s$^{-1}$. These molecules were also reported by \citet{corby2015} at a velocity around 0~km\,s$^{-1}$, but they detected only one velocity component due to their poor spectral resolution (6--10~km\,s$^{-1}$). While \citet{corby2015} did not detect complex molecules outside the GC, we report here the detection of CH$_3$OH and CH$_3$CN in a diffuse cloud in the Scutum arm. Figure~\ref{ntots} shows that the non-detection of the other complex species in the Scutum arm does not necessarily reflect a different chemical composition compared to the diffuse GC clouds but may simply be due to a lack of sensitivity.

Methanol's high abundance (10$^{-9}$--10$^{-8}$) in translucent (and hence perhaps also in diffuse) molecular clouds could be driven by UV photodesorption from dust grains \citep{turner1998}. This process would not be effective in denser parts of molecular clouds which are mostly shielded against UV radiation. More recently, a process of grain-surface formation and immediate reactive desorption was found to be an important source of gas-phase methanol \citep{garrod2007}. We find methanol abundances relative to H$_2$ on the order of 10$^{-7}$ for the diffuse GC and Scutum-arm clouds (Table~\ref{hi_h2}), which is surprisingly high, one order of magnitude higher than in translucent clouds \citep{turner1998} and dense prestellar cores such as L1544 \citep[$6 \times 10^{-9}$][]{vastel2014}. Although we cannot rule out that the higher UV radiation field in diffuse clouds may enhance the abundance of methanol, it may be that the abundances we obtain are overestimated due to an underestimate of the H$_2$ column density if the clouds are clumpy on a scale intermediate between the resolution of the Herschel HF data (40$\arcsec$) and our ALMA survey (1.6$\arcsec$). CH$_3$CN can be formed effectively in the gas phase without requiring grain-surface formation \citep{turner1999,bergner2017}. CH$_3$CN is enhanced relative to methanol in the Scutum-arm diffuse cloud in comparison to the GC diffuse clouds. This is also the case for the Horsehead PDR. The high abundance of CH$_3$CHO in translucent clouds cannot be explained by gas-phase chemistry only and represents maybe the border to grain chemistry \citep{turner1999}. This is most likely also true for the diffuse GC clouds, where we detect CH$_3$CHO.

The GC and Scutum-arm clouds have a molecular hydrogen fraction between 0.4 and 0.7, implying that they are indeed diffuse molecular clouds \citep{snow2006}. However, the \ion{H}{i} and H$_2$ column densities of the diffuse GC clouds are one order of magnitude higher than those of the Scutum-arm cloud while their molecular fraction is lower (Table~\ref{hi_h2}), which appears to be contradictory. This may be an indication that the molecular parts of these clouds are clumpy, and in turn that the molecular fraction of the regions where H$_2$ resides is underestimated. This argument supports our previous hypothesis that the methanol abundances derived above are overestimated due to an underestimate of the H$_2$ column densities.

\citet{price2003} introduced a theory, further tested by \citet{garrod2005,garrod2006}, in which interstellar clouds, or clumps within them, cycle between dense and diffuse conditions, with the chemistry of the diffuse stages enriched by the survival of molecules formed under denser conditions. Their models showed abundance enhancements of several orders of magnitude for some of the larger molecules. COMs in diffuse clouds may thus be the remnants of a previous denser phase in this cyclic process.

\section{Conclusions}\label{sect_conc}
We report the detection of the (complex) organic molecules CH$_3$OH, CH$_3$CN, CH$_3$CHO, HC$_3$N, and NH$_2$CHO in diffuse GC clouds along the line of sight to Sgr\,B2(N) and CH$_3$OH and CH$_3$CN in a diffuse cloud of the Scutum arm. The chemical composition of one of the diffuse GC clouds is found to be generally similar to the one of the diffuse/translucent medium of the $z=0.89$ spiral galaxy in front of PKS~1830--211, suggesting that the chemical processes leading to chemical complexity have remained similar since $z=0.89$. Some differences are seen, however: HC$_3$N appears to be much more abundant relative to methanol in this diffuse GC cloud compared to the spiral galaxy, while CH$_3$CN is enhanced in the Scutum-arm cloud, like in the Horsehead PDR. The presence of COMs in diffuse molecular clouds may result from a cyclical interstellar process of cloud contraction and expansion between diffuse and dense states \citep{price2003, garrod2005,garrod2006}. 

\begin{acknowledgements} 
This paper makes use of the following ALMA data: ADS/JAO.ALMA\#2011.0.00017.S, ADS/JAO.ALMA\#2012.1.00012.S. ALMA is a partnership of ESO (representing its member states), NSF (USA) and NINS (Japan), together with NRC (Canada), NSC and ASIAA (Taiwan), and KASI (Republic of Korea), in cooperation with the Republic of Chile. The Joint ALMA Observatory is operated by ESO, AUI/NRAO and NAOJ. The interferometric data are available in the ALMA archive at https://almascience.eso.org/aq/. 
\end{acknowledgements}

%
\bibliographystyle{aa} 
\bibliography{references} 

\begin{appendix}

\section{Additional figures}
  \begin{figure}[h]
\centering
  \resizebox{\hsize}{!}{\includegraphics[trim = 2cm 1.5cm 6.5cm 3.5cm, clip=True]{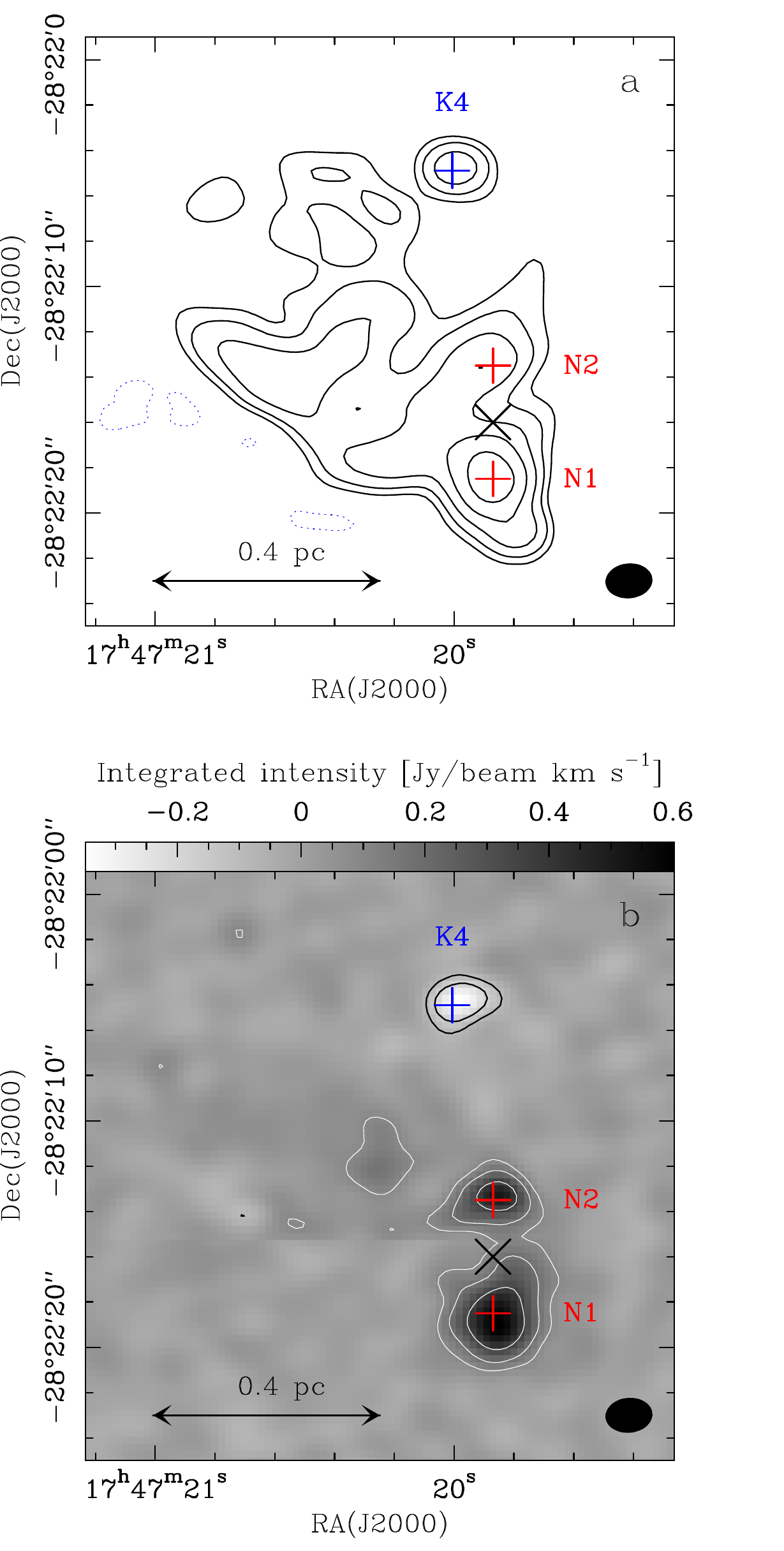}}
\caption{\textbf{a} ALMA continuum map of the Sgr B2(N) region at 85~GHz. The level of the first contour (positive as black solid line and negative as dashed blue line) is $5\sigma$, with $\sigma$ the rms noise level of 5.4~mJy/beam. The contour step doubles at each contour. \textbf{b} Integrated intensity map of the NH$_2$CHO absorption of the GC diffuse clouds ($v_\mathrm{LSR}=-8.9$--$6.6$~km~s$^{-1}$) at 84.5~GHz. The first contour level (positive in white and negative in black) is $5\sigma$, with $\sigma=17.7$~mJy/beam~km~s$^{-1}$. The contour step doubles at each contour. The grey-scale is indicated at the top. The continuum emission was subtracted. In both panels, the filled ellipse shows the synthesised beam ($2.1\arcsec\times1.5\arcsec$), the black cross indicates the phase centre, the red crosses the positions of the hot cores Sgr~B2(N1) and Sgr~B2(N2), and the blue cross the position of the ultracompact \ion{H}{ii} region K4.} 
\label{cont}
\end{figure}

\section{Additional tables}
  \begin{table}[h]
\caption{Rest frequencies ($\nu_0$), upper energy levels ($E_\mathrm{up}$) and Einstein coefficients for spontaneous emission ($A_\mathrm{u,l}$) from upper level $u$ to lower level $l$ of the molecular transitions investigated in this work.}
\label{transitions}
\centering
\begin{tabular}{l c r r c}       
\hline               
Molecule & Transition & \multicolumn{1}{c}{$\nu_0$} & \multicolumn{1}{c}{$E_\mathrm{up}$} & $A_\mathrm{u,l}$\\
 & & \multicolumn{1}{c}{[MHz]} & \multicolumn{1}{c}{[K]} & [s$^{-1}$]\\
\hline
\hline               
H$^{13}$CO$^+$ & $1-0$ & 86754.288 & 4.2 & $3.85\times10^{-5}$\\  
c-C$_3$H$_2$-o & $2_\mathrm{1,2}-1_\mathrm{0,1}$ & 85338.894 & 4.1 & $2.32\times10^{-5}$\\
CH$_3$OH & $2_\mathrm{0,1,0}-1_\mathrm{0,1,0}$ & 96741.371 & 7.0 & $3.41\times10^{-6}$\\
$^{13}$CH$_3$OH & $2_\mathrm{0,2,0}-1_\mathrm{0,1,0}$ & 94407.129 & 6.8 & $3.17\times10^{-6}$\\
CH$_3$CN & $5_\mathrm{0,0}-4_\mathrm{0,0}$ & 91987.088 & 13.2 & $6.33\times10^{-5}$\\
 & $6_\mathrm{0,0}-5_\mathrm{0,0}$ & 110383.500 & 18.5 & $1.11\times10^{-4}$\\
CH$_3$CHO & $2_\mathrm{1,2,0}-1_\mathrm{0,1,0}$ & 84219.749 & 5.0 & $3.38\times10^{-6}$\\
 & $5_\mathrm{0,5,0}-4_\mathrm{0,4,0}$ & 95963.459 & 13.8 & $2.95\times10^{-5}$\\
 & $3_\mathrm{1,3,1}-2_\mathrm{0,2,2}$ & 101343.441 & 7.7 & $3.91\times10^{-6}$\\
HC$_3$N & $10-9$ & 90979.023 & 24.0 & $5.81\times10^{-5}$\\
 & $11-10$ & 100076.392 & 28.8 & $7.77\times10^{-5}$\\
NH$_2$CHO & $4_\mathrm{0,4}-3_\mathrm{0,3}$ & 84542.330 & 10.2 & $4.09\times10^{-5}$\\
 & $4_\mathrm{1,3}-3_\mathrm{1,2}$ & 87848.874 & 13.5 & $4.30\times10^{-5}$\\
 & $5_\mathrm{1,5}-4_\mathrm{1,4}$ & 102064.267 & 17.7 & $7.06\times10^{-5}$\\
 & $5_\mathrm{0,5}-4_\mathrm{0,4}$ & 105464.219 & 15.2 & $8.11\times10^{-5}$\\
\hline                      
\end{tabular}
\end{table}

\begin{table}[h]
\caption{\ion{H}{i} and H$_2$ column densities and molecular hydrogen fraction $f_\mathrm{H_2}$ determined by \citet{winkel2017}, and abundance of methanol relative to H$_2$.}
\label{hi_h2}
\centering
\begin{tabular}{c c c c c}       
\hline               
$v_\mathrm{LSR}$ & $N_\mathrm{\ion{H}{i}}$ & $N_\mathrm{H_2}$ & $f_\mathrm{H_2}$ & $\frac{N_\mathrm{CH_3OH}}{N_\mathrm{H_2}}$ \\

[km\,s$^{-1}$] & [cm$^{-2}$] & [cm$^{-2}$] & & \\
\hline\hline              
26.7 & $4.7\times 10^{20}$ & $5.3\times 10^{20}$ & 0.70 & $1.2\times 10^{-7}$\\
9.4 & $4.7\times 10^{21}$ & $4.7\times 10^{21}$ & 0.55 & $8.1\times 10^{-8}$\\
3.4 & $6.9\times 10^{21}$ & $1.9\times 10^{21}$ & 0.36 & $2.0\times 10^{-7}$\\
$-2.8$ & $6.3\times 10^{21}$ & $1.7\times 10^{21}$ & 0.40 & $1.2\times 10^{-7}$\\
\hline                      
\end{tabular}
\end{table}

\end{appendix}

\end{document}